\documentclass[aps,eqsecnum,nofootinbib]{revtex4}
\usepackage{amsmath,amsfonts,amssymb,bm}

\newcommand{\be}{\begin{equation}}
\newcommand{\ee}{\end{equation}}
\newcommand{\bea}{\begin{eqnarray}}
\newcommand{\eea}{\end{eqnarray}}

\newcommand{\nq}{\mathfrak{q}}
\newcommand{\nw}{\mathfrak{w}}

\newcommand{\tpi}{\tau_{\pi}}

\newcommand{\eq}[1]{Eq.~(\ref{eq:#1})}

\newcommand{\cR}{{\cal R}}

\begin{document}

%\begin{CJK*}{SJIS}{} 

%%%%%%%   Title Page   %%%%%%%
%
\title{
A note on causal hydrodynamics for M-theory branes
}
\author{Makoto Natsuume}
\email{makoto.natsuume@kek.jp}
\affiliation{Theory Division, Institute of Particle and Nuclear Studies, \\
KEK, High Energy Accelerator Research Organization, Tsukuba, Ibaraki, 305-0801, Japan}
\author{Takashi Okamura}
\email{tokamura@kwansei.ac.jp}
\affiliation{Department of Physics, Kwansei Gakuin University,
Sanda, Hyogo, 669-1337, Japan}
\date{\today}
\begin{abstract}
We obtain new transport coefficients of causal hydrodynamics for the M2 and the M5-branes using a Kubo-like formula proposed by Baier, Romatschke, Son, Starinets, and Stephanov (arXiv:0712.2451 [hep-th]). The relaxation times agree with the ones obtained from the ``sound mode" in our paper (arXiv:0712.2916 [hep-th]).
\end{abstract}
%
%\pacs{11.25.Tq, 12.38.Mh} %11.25.-w, 11.25.Uv, 
% KEK-TH-1221

\maketitle
%\end{CJK*}{SJIS}{}

%%%%%%%%%
\section{Introduction}
%%%%%%%%%

The AdS/CFT duality is a powerful tool to study hydrodynamics of gauge theory plasmas (See Ref.~\cite{Natsuume:2007qq} for a review). However, standard hydrodynamics (first-order formalism) has severe problems such as acausality. 
One can restore causality by introducing a new set of transport coefficients. The resulting theory is known as ``causal hydrodynamics"  or  ``second-order formalism." At present, there is no unique formalism for causal hydrodynamics, but probably the most used formalism is the ``Israel-Stewart theory" \cite{Israel:1976tn,Israel:1979wp}. 

The Israel-Stewart theory has been widely discussed in the context of heavy-ion collisions. Recently, a number of papers appeared which study the causal hydrodynamics of gauge theory plasmas using the AdS/CFT duality \cite{Benincasa:2007tp}-\cite{Natsuume:2007tz}.%
\footnote{See also Ref.~\cite{Heller:2007qt} for an early work and Ref.~\cite{causal_review} for a review.}
Some of the main lessons drawn from these works are,
\begin{enumerate}
\item One should not take the Israel-Stewart theory too literally; it should be really considered as an effective theory.
%Gauge theory plasmas do not fit into the framework of the Israel-Stewart theory. 
For example, a single transport coefficient $\tpi$ appears both in the ``shear mode" and in the ``sound mode." However, if one naively uses the Israel-Stewart theory to interpret the AdS/CFT results, two different values arise.
%but two different values arise from the AdS/CFT duality 
%This strongly suggests that the Israel-Steart theory is not sufficient to describe gauge theory plasmas or it is not the most general effective theory. 
Reference~\cite{Baier:2007ix} argues that the shear mode result is unreliable due to the corrections from ``third-order" hydrodynamics. Namely, the shear mode case exceeds the validity of the effective theory.
References~\cite{Baier:2007ix,Natsuume:2007ty} confirmed this for SAdS$_5$,%
\footnote{The five-dimensional Schwarzschild-AdS black hole, which is dual to the ${\cal N}=4$ SYM.}
and Ref.~\cite{Natsuume:2007ty} confirmed this for SAdS$_4$ and SAdS$_7$ as well.

\item Reference~\cite{Baier:2007ix} proposed a Kubo-like formula for conformal theories to obtain $\tpi$ from the ``tensor mode." The formula yields the value of $\tpi$ which appears in the sound mode for SAdS$_5$. 

% v2
\item In addition to the coefficients introduced by Israel and Stewart, Ref.~\cite{Baier:2007ix} also introduced new coefficients $\kappa$ and $\lambda_{1,2,3}$, which are necessary for consistency. Although $\kappa$ contributes to the Kubo-like formula, for small perturbations and for flat space, which were the main subject of Refs.~\cite{Baier:2007ix,Natsuume:2007ty}, they do not contribute to the energy-momentum tensor, and their theory reduces to the Israel-Stewart theory. 
\end{enumerate}
In this paper, we use the Kubo-like formula for SAdS$_{4,7}$ to obtain the  coefficient $\tpi$. The values of $\tpi$ agree with the one obtained from the sound mode in Ref.~\cite{Natsuume:2007ty}. We also obtain the coefficient $\kappa$ for SAdS$_7$, which has been evaluated only for SAdS$_5$. It is desirable to compute $\kappa$ in various theories to uncover any generic behavior as the ratio of the shear viscosity to the entropy density.

%%%%%%%%%
\section{Tensor mode computation for SAdS$_7$}
%%%%%%%%%

The ${\rm SAdS}_{p+2}$ metric is given by
\be
ds_{p+2}^2 = 
f (-hdt^2+d\vec{x}_p^2) +  \frac{dr^2}{fh}~,
\label{eq:sads_metric}
\ee
where 
\bea
f &=& \left( \frac{r}{R} \right)^2~, \\
h &=& 1-\left( \frac{r_0}{r} \right)^{p+1}~,
\eea  
and the temperature is given by
\be
2\pi T=\frac{p+1}{2}\frac{r_0}{R^2}~.
\ee
It is convenient to use  a new radial coordinate  $u$ ($u:=r_0/r$ for even $p$, and $u:=r_0^2/r^2$ for odd $p$). From the point of view of string theory, the interesting cases are $p=2,3$, and 5. The $p=3$ case corresponds to the D3-brane which is dual to the ${\cal N}=4$ SYM. The $p=2$ and 5 cases correspond to the M2 and the M5-branes in the 11-dimensional supergravity, respectively. 

According to the standard AdS/CFT dictionary, the bulk gravitational perturbations act as the source for the stress-energy tensor of the dual boundary theory. Since the main object in hydrodynamics is the stress-energy tensor, our aim is to solve the bulk gravitational field equations. 
Consider the bulk perturbations of a $p$-brane which take the form 
$
h_{\mu\nu}=H_{\mu\nu}(r) \, e^{-iwt + iqz},
$
where $z:=x^p$. The perturbations can be decomposed by the little group $SO(p-1)$ acting on $x^i (i = 1, \cdots, p-1)$. The gravitational perturbations are decomposed as the tensor mode, the vector mode (``shear mode"), and the scalar mode (``sound mode"). 
Similar decomposition is employed in hydrodynamics. 
These modes are studied in the framework of causal hydrodynamics by various references:%
\footnote{
In the context of standard hydrodynamics, the $p=3$ case has been first studied in Refs.~\cite{Policastro:2002se,Policastro:2002tn}, and the $p=2,5$ cases have been studied in Refs.~\cite{Herzog:2002fn,Herzog:2003ke}. }
% JHEP v2?
%For $p=3$, the tensor mode is studied in Ref.~\cite{Baier:2007ix}; the shear mode is studied in Refs.~\cite{Baier:2007ix,Bhattacharyya:2007jc,Natsuume:2007ty} (See also Ref.~\cite{Heller:2007qt}); and the sound mode is studied in Refs.~\cite{Baier:2007ix,Bhattacharyya:2007jc,Natsuume:2007ty}. For $p=2,5$, the shear mode and the sound mode are studied in Ref.~\cite{Natsuume:2007ty}.%
\begin{center}
\begin{tabular}{|c|l|l|}
\hline
SAdS$_5$
	& tensor mode & Ref.~\cite{Baier:2007ix}\\
	& shear mode & Ref.~\cite{Baier:2007ix,Natsuume:2007ty} (also Refs.~\cite{Heller:2007qt,Bhattacharyya:2007jc})\\
	& sound mode & Ref.~\cite{Baier:2007ix,Natsuume:2007ty} (also Ref.~\cite{Bhattacharyya:2007jc})\\
\hline
SAdS$_4$ and SAdS$_7$
 	& tensor mode & $-$ \\
	& shear mode & Ref.~\cite{Natsuume:2007ty}\\
	& sound mode & Ref.~\cite{Natsuume:2007ty}\\
\hline
\end{tabular}
\end{center}
So, the tensor mode computations for $p=2, 5$ are missing: we presently compute the so far missing tensor mode for these cases.%We supply this missing link.%
\footnote{The word ``tensor mode" is not really appropriate for $p=2$ since the tensor mode exists for branes with $p \geq 3$. Nevertheless, we keep using this terminology since there is a Kubo-like formula even for $p=2$. See \eq{BRS3_formula3}. Even though it is really the shear mode, the shear mode computation presented in Ref.~\cite{Natsuume:2007ty} is based on quasinormal frequencies, and it is different from the one here.}

The tensor mode can determine $\tpi$ through a Kubo-like formula (for conformal theories) \cite{Baier:2007ix}:
\be
G_{xy,xy}^R = P -i \eta w + \eta \tpi w^2 - \frac{\kappa}{2} [(p-2)w^2+q^2]~,
\qquad (p \geq 3),
\label{eq:BRS3_formula} 
\ee
where
\begin{center}
\begin{tabular}{cl}
$G_{xy,xy}^R$ &: retarded Green's function for the tensor mode $h_{xy}(t,z)$~,\\
$P$ &: pressure, \\
$\eta$ &: shear viscosity, \\
$\tpi$ &: relaxation time for the shear viscous stress, \\
$\kappa$ &: a new coefficient introduced in Ref.~\cite{Baier:2007ix}.
\end{tabular}
\end{center}For $p=2$, one should consider the perturbation of the form $h_{xz}(t)$ instead:
\be
G_{xz,xz}^R = P -i \eta w + \eta \tpi w^2 ~,
\qquad (p = 2).
\label{eq:BRS3_formula3}
\ee
Using these formulae, we obtain $\tpi$ for SAdS$_{4,7}$ and $\kappa$ for SAdS$_{7}$.

The total action for SAdS$_{p+2}$ consists of
\be
S_{\rm total} = \frac{{\rm vol}({\bf S}^{9-p})}{16 \pi G_{11}}
\big\{ S_{\rm bulk} + S_{\rm GH} + S_{\rm CT} \big\}~,
\label{eq:total_action}
\ee
where ${\rm vol}({\bf S}^{9-p})$ is the volume of the compactified ${\bf S}^{9-p}$ with radii $(2R, R/2)$ for $p=(2, 5)$, respectively. The 11-dimensional Newton constant $G_{11}$ is written as 
\be
\frac{1}{16\pi G_{11}} = \frac{N_c^{3/2}}{\sqrt{2} \pi^5 (2R)^9} \quad (p=2)~, \qquad
\frac{1}{16\pi G_{11}} = \frac{2N_c^3}{\pi^5R^9} \quad (p=5)~.
\ee
Also, $S_{\rm bulk}$, $S_{\rm GH}$, and $S_{\rm CT}$ are the bulk action, the familiar Gibbons-Hawking action, and the counterterm action which cancels divergences, respectively:
\bea
S_{\rm bulk} &=& \int d^{p+2}x\, \sqrt{-g} \left\{ {\bf R} + \frac{p(p+1)}{R^2} \right\}~, \\
S_{\rm GH} &=& 2\int d^{p+1}x\, \sqrt{-\gamma}\, K~, \\
S_{\rm CT} &=& -\int  d^{p+1}x\, \sqrt{-\gamma} 
\left\{ 
\frac{2p}{R} 
+ \frac{R}{p-1}\cR - \frac{R^3}{(p-3)(p-1)^2} \left(\cR^{\mu\nu}\cR_{\mu\nu} - \frac{p+1}{4p}\cR^2\right)
%+ \cdots
\right\}. 
\eea
Here, ${\bf R}$ is the bulk Ricci scalar, $\gamma_{ab}$ is the boundary metric restricted to $u=0$, and $K$ is the trace of the extrinsic curvature of the boundary: $K = -\partial_u\sqrt{-\gamma}/(N \sqrt{-\gamma})$, where $N$ is the lapse function given by $N^2 := 1/g^{uu}$. Also, $\cR_{\mu\nu}$ and $\cR$ are the Ricci tensor and the Ricci scalar for the boundary metric $\gamma_{ab}$, respectively. 

These counterterms are sufficient to cancel power-law divergences for $p \leq 5$. For $p=5$, also other counterterms exist which cancel logarithmic divergences  (See, {\it e.g.},  Ref.~\cite{Papadimitriou:2004ap}), but these terms are not necessary for our computations. In addition, the $O(\cR^2)$ terms are not actually necessary either for our computations because they are $O(w^4,w^2q^2,q^4)$.
% v2

In this section, we consider the $p=5$ case. As usual, choose the gauge where $h_{xy}\neq0$ 
with the other $h_{\mu\nu}=0$. 
Introducing the variable $h^x_{~y} =:\phi_{w,q}\, e^{-iw t + i q z} $, $\phi_{w,q}$ satisfies the equation for a minimally-coupled scalar field:
\be
\phi_{\nw,\nq}'' + \left( \frac{h'}{h} - \frac{2}{u} \right)\phi_{\nw,\nq}' + \frac{9}{4} \frac{1}{u h^2} (\nw^2-h \nq^2) \phi_{\nw,\nq} = 0~,
\ee
where $':=\partial_u$, $\nw=w/(2\pi T)$, and $\nq=q/(2\pi T).$
Incorporating the ``incoming wave" boundary condition at the horizon $u=1$, the solution is given by
\be
\phi_{\nw,\nq}(u)=Ch^{-i\nw/2} F(u)~,
\ee
where $F$ is a regular function whose form can be obtained perturbatively as a double series in $\nw$ and $\nq$. An integration constant in $F$ is fixed by requiring the solution to be regular at the horizon, and the constant $C$ is fixed so that $\phi_{\nw,\nq}(0)=1$.  The full solution is a rather cumbersome expression, 
so we do not write it here, 
% JHEP v2?
%which can be found in Appendix,
 but it is enough to keep the terms of $O(u^3)$ to obtain $\tpi$ from the Kubo formula:
\be
\phi_{\nw,\nq} \sim 
1+ \frac{1}{2} i\nw u^3 
+\frac{3}{8} u(-3+u^2) \nq^2
+\frac{u}{24} \{ 27+(-9+9 \ln 3+\sqrt{3}\pi)u^2 \}  \nw^2 + \cdots~.
\label{eq:tensor_solution}
\ee

Rewriting the bulk action as a boundary action by the bulk equation of motion, and evaluating the total action (\ref{eq:total_action}), we find the following boundary action:
\be
S_{\rm total} = \left. -\frac{\pi^3}{3}\left(\frac{2}{3}\right)^6 N_c^3 T^6
\left[ V_6- \frac{1}{u^2} \phi_{-\nw,-\nq} \phi_{\nw,\nq}' 
+ \left\{ \frac{1}{2} -\frac{9(\nq^2-\nw^2)}{8u^2} \right\} \phi_{-\nw,-\nq}\phi_{\nw,\nq} + \cdots \right] \right|_{u=0}~.
\ee
The fluctuation-independent part gives the pressure. The part quadratic in fluctuations gives the two-point function by the recipe of Ref.~\cite{Son:2002sd}. 
Substituting the solution (\ref{eq:tensor_solution}), we find
\be
G_{xy,xy}^R = \frac{2\pi^3}{3}\left(\frac{2}{3}\right)^6 N_c^3 T^6 
\left\{ -\frac{3}{2} i \nw 
- \frac{9}{8} \nq^2 -\frac{1}{8} (-9+9 \ln 3+\sqrt{3}\pi) \nw^2 + \frac{1}{2} + \cdots
\right\}~.
\label{eq:retarded7}
\ee
Comparing \eq{retarded7} to the Kubo formula (\ref{eq:BRS3_formula}), we obtain the familiar results of the pressure and the viscosity \cite{Herzog:2002fn} as well as the two parameters of the causal hydrodynamics:
\be
P = \frac{\pi^3}{3}\left(\frac{2}{3}\right)^6 N_c^3 T^6~, \quad
\eta = \frac{\pi^2}{2}\left(\frac{2}{3}\right)^6 N_c^3 T^5~, \quad
\tpi = \frac{36-(9\ln3+\sqrt{3}\pi)}{24\pi T}~, \quad
\kappa = \frac{3\eta}{4\pi T}~.
\ee
The value of $\tpi$ is the same as the one obtained from the sound mode but disagrees with the one obtained from the shear mode \cite{Natsuume:2007ty}. This observation was made for the ${\cal N}=4$ SYM \cite{Baier:2007ix}, and we found that this is the case for SAdS$_7$ as well.

For the ${\cal N}=4$ SYM, $\kappa=\eta/(\pi T)$. Thus, the ratio of $\kappa$ to $\eta$ is not universal. But at present, it is not even clear if $\kappa$ must be proportional to $\eta$.  %Also, $\tpi$ obtained from the AdS/CFT duality is roughly the same order of magnitude as $\tpi$ obtained from the kinetic theory \cite{Natsuume:2007ty}. But if $\kappa$ is proportional to $\eta$ in general, this is not true for $\kappa$ since $\eta$ strongly depends on the coupling constant. 
It would be interesting to evaluate $\kappa$ in perturbative QCD; if $\kappa$ is proportional to $\eta$, the perturbative estimate of $\kappa$ must be much larger than the AdS/CFT value since $\eta$ becomes large at weak coupling.

%%%%%%%%%
\section{``Tensor mode" computation for SAdS$_4$}
%%%%%%%%%

The computation for SAdS$_4$ is similar, so we will be brief.
Again, $h^x_{~z} =: \phi_{w}\, e^{-iw t}$ satisfies the equation for a minimally-coupled scalar field:
\be
\phi_{\nw}'' + \left( \frac{h'}{h} - \frac{2}{u} \right)\phi_{\nw}' + \frac{9\nw^2}{4h^2} \phi_{\nw} = 0~.
\ee
Since $h_{xz}$ is really a field in the shear mode, it normally couples to another filed in the shear mode $h_{xt}$ (in the gauge $h_{xu}=0$), but they decouple in the $\nq=0$ limit. Solving the differential equation, and keeping the terms of $O(u^3)$, we get
\be
\phi_{\nw} \sim 
1+ \frac{1}{2} i\nw u^3 
+\frac{u^2}{24} \{ 27-(18-9 \ln 3+\sqrt{3}\pi)u \}  \nw^2 + \cdots~.
\label{eq:tensor_solution4}
\ee

Evaluating the total action (\ref{eq:total_action}), we find the following boundary action:
\be
S_{\rm total} = \left. -\frac{\pi^2}{12\sqrt{2}}\left(\frac{4}{3}\right)^3 N_c^{3/2} T^3
\left[ V_3- \frac{1}{2u^2} \phi_{-\nw} \phi_{\nw}' 
+ \left\{ \frac{1}{2} +\frac{9\nw^2}{8u^2} \right\} \phi_{-\nw}\phi_{\nw} + \cdots \right] \right|_{u=0}~.
\ee
Substituting the solution (\ref{eq:tensor_solution4}), we find
\be
G_{xz,xz}^R = \frac{\pi^2}{6\sqrt{2}}\left(\frac{4}{3}\right)^3 N_c^{3/2} T^3 
\left\{ -\frac{3}{4} i \nw 
+\frac{1}{16} (18-9 \ln 3+\sqrt{3}\pi) \nw^2 + \frac{1}{2} + \cdots
\right\}~.
\label{eq:retarded4}
\ee
Comparing \eq{retarded4} to the Kubo formula (\ref{eq:BRS3_formula3}), we obtain 
\be
P = \frac{\pi^2}{12\sqrt{2}} \left(\frac{4}{3}\right)^3 N_c^{3/2} T^3~, \quad
\eta = \frac{\pi}{16\sqrt{2}} \left(\frac{4}{3}\right)^3 N_c^{3/2} T^2~, \quad
\tpi = \frac{18-(9\ln3-\sqrt{3}\pi)}{24\pi T}~.
\ee
Again, the value of $\tpi$ is the same as the one obtained from the sound mode  \cite{Natsuume:2007ty}. The value of $\eta$ also agrees with the one by the other methods \cite{Herzog:2002fn}, which gives a consistency check of the Kubo formula (\ref{eq:BRS3_formula3}) even for $p=2$.

%%%%%%%%%
\section{Discussion}
%%%%%%%%%

% v2
The M-theory branes considered in this paper has the $d=10$ interpretations: the M5-brane as the D4-brane, and the M2-brane as the D1-brane. Thus, the D4-brane should have the Green's function $G_{xy,xy}^R$ with the same $(\nw,\nq)$-dependence as in \eq{retarded7}. Similarly, a Green's function for the D1-brane should take the form of \eq{retarded4}. However, these D-branes are non-conformal theories; the Kubo-like formula (\ref{eq:BRS3_formula}) applies only to conformal theories. Following the construction of Ref.~\cite{Baier:2007ix}, one can see that an additional transport coefficient (say $\kappa'$) should appear in the Kubo-like formula for nonconformal theories, so one can no longer determine these coefficients separately. 

\acknowledgments
We would like to thank Harvey Meyer for his various comments on the manuscript. We would also like to thank the anonymous referee who pointed us an error in \eq{tensor_solution} on the draft submitted to Progress of Theoretical Physics. 

%\footnotesize

\end{document}